\magnification=\magstep1 
\baselineskip=14 pt
\hsize=5 in
\vsize=7.3 in
\pageno=1

\vskip 1,8 cm
\centerline{\bf ROTATION OPERATOR VS PARTICLE CREATION}
\centerline{\bf IN A CURVED SPACE TIME}
\vskip 1,8 cm 
\centerline{\bf Carlos E. Laciana}
\centerline{\sl Instituto de Astronom\'{\i}a y F\'{\i}sica del Espacio} 
\centerline{\sl Casilla de Correo 67 - Sucursal 28, 1428 Buenos Aires, 
Argentina}
\centerline{\sl E-mail: laciana@iafe.uba.ar}
\vskip 1,8 cm

\vskip 0,2cm 
 Taking into account a neutral massive scalar field minimally coupled to  
gravity, in a Robertson-Walker metric, it is shown that when the final state 
is connected with the initial one by means of a Bogoliubov transformation, 
which does not include the single-mode rotation operator, the mean value of 
created particles  is  conserved.    When  the rotation operator is considered,
 it is still possible to use the approach of single-mode squeezed operators 
and get the entropy as the logarithm of the created particles. 
\vskip 2,5 cm
\vfill\eject
 
 In the  work  by  Gasperini  and  Giovannini  [1] it is shown that the large
amount of entropy of our present universe can be explained by the mechanism 
of cosmological pair creation from the vacuum, as a consecuence  
a background of squeezed 
cosmic gravitons is produced.  In our case we will study the contribution of 
a neutral massive scalar field to the entropy, without any graviton present, 
in a classical curved background. In order to do that we will 
use  the formalism  introduced  by  Parker  [2] which let us define  
the general  time  dependent
Bogoliubov transformation in a Robertson-Walker metric. In that reference 
a consistent method to obtain the annihilation-creation operators at each time 
is given.    In  this  work  we  will  show  that  the general time depending
Bogoliubov transformation, introduced in ref. [2], transforms the  
``in" vacuum to a squeezed state. 

 Ref.[1] is based on the approach of ref.[3] where  a linearized 
graviton theory is used. In that case the graviton modes 
satisfy the Klein-Gordon 
equation corresponding to a massless scalar field minimally coupled to the 
gravitational background.    In  the  present  case  we use a massive scalar
field. In both cases the procedure is similar. In ref.[4] we can see that 
still in the massless case the particle creation number is different from zero, 
due to the  kind  of  coupling  (with  the  conformal  coupling  the particle
creation is zero [5]).   

 Let us now rewrite the expresions of ref.[2] that we need in the following. 
The minimally coupled scalar field satisfies the field equation 
 
$$({{\bigtriangledown}_{\mu} {\partial}^{\mu}}  +  {m^2}){\varphi}  =  0\eqno
(1)$$
\vskip 0,2 cm
(with $\mu = 0,1,2,3$). 
\vskip 0,2 cm

 For an unbounded space-time, as we use, a convenient technical method is the 
discretization in the momentum space, which is implemented as in ref.  [2] by 
means of  the  boundary condition $\varphi ({\bf x} + {\bf n}L)=
\varphi ({\bf x},t)$, where ${\bf n}$ is a vector with integer Cartesian 
components and $L$  a  longitude  which  goes  to  infinity at the end of the
calculation (in our case it is not necessary because the results used do not 
depend on $L$). Then a general solution of eq. (1) can be expanded in the 
form 
$${\varphi ({\bf x}, t)} = {\sum _{\bf k}}[{a_{\bf k}}(t) 
{\phi _{\bf k}}(x) + {{a^{\dagger}}_{\bf k}}(t) 
{{\! \phi ^{\ast}}_{\bf k}}(x)]\eqno (2)$$ 
\vskip 0,2 cm
where  
$${{\phi _{\bf k}}(x)} = {1\over {(L a(t))^{3/2} {\sqrt {2W}}}} \exp {i(
{{\bf k}{\bf x}} - {\int _{t_{0}}}^{t} W(k, {t^{\prime}}) dt^{\prime})}$$ 
\vskip 0,2 cm
whith $W$ a real function of $k=|{\bf k}|$ and $t$.  The operators 
$a_{\bf k}(t)$ and $a_{{\bf k}^{\prime}}(t)$ are time dependent and satisfy the 
commutation relations 
\vskip 0,2 cm 
$$[a_{\bf k}(t), a_{\bf k^{\prime}}(t)] = 0,\ \ \ 
[{{\! a^{\dagger}}_{\bf k}}(t), 
{{\! a^{\dagger}}_{\bf k^{\prime}}}(t)] = 0,\ \ \ 
[{a_{\bf k}}(t), {{\! a^{\dagger}}_{\bf k^{\prime}}}(t)] = 
\delta _{{\bf k}, {{\bf k}^{\prime}}}\eqno (3)$$
\vskip 0,2cm
The operators act on the vacuum $|0,t>$ in the form 
\vskip 0,2 cm
$${a_{\bf k}}(t)|0,t>=0,\  \  \  {{\!  a^{\dagger}}_{\bf k}}(t)|0,t>=|{1_{\bf
k}},t>$$
\vskip 0,2 cm
Moreover we define the operators
$${A_{\bf k}}:= {a_{\bf k}}(t=t_{1})\ \ \ with\ \ t\ge t_{1}\ge t_{0}$$
$${A_{\bf k}}^{\dagger}:={{a_{\bf k}}^{\dagger}}(t=t_{1})$$ 
Also we define the vacuum $|0>:=|0,t=t_{1}>$ on which the operators 
defined before act. These operators are related with the time dependent 
ones by the following Bogoliubov transformation 
\vskip 0,2 cm
 
$${a_{\bf k}}(t) = {{e^{i\gamma_{\alpha}({\bf k},t)}}\cosh r({\bf
k},t) }A_{\bf k}  + 
{{e^{i\gamma_{\beta}({\bf  k},t)}}\sinh r({\bf
k},t) }{A^{\dagger}}_{-{\bf k}}\eqno (4a)$$
\vskip 0,2 cm
$${a_{-{\bf k}}}^{\dagger}(t) = {{e^{-i\gamma_{\beta}({\bf  k},t)}}
\sinh r({\bf k},t) }A_{\bf k} + 
{{e^{-i\gamma_{\alpha}({\bf k},t)}}\cosh r({\bf
k},t) }{A^{\dagger}}_{-{\bf k}}   \eqno (4b)$$
\vskip 0,2 cm
where the functions $\gamma_{\alpha}$ and $r$ satisfy the 
obvious conditions 
$${\gamma_{\alpha}}({\bf k},{t_{1}}) = 0.$$
$${r({\bf k},{t_{1}}) = 0.}$$
\vskip 0,2 cm
Replacing eqs.(4)  in  eq.    (2)  and  all of that in eq.  (1) we obtain the
following system of equations (see ref. [6])
\vskip 0,2 cm
$$(1 + {\tanh r}\cos \Gamma)M + 2W{{\dot \gamma}_{\alpha}} 
= 0\eqno (5a)$$
\vskip 0,2 cm
$$M{\sin \Gamma} + 2W{\dot r} = 0\eqno (5b)$$
\vskip 0,2 cm
with $\Gamma:= {\gamma_{\alpha}} + {\gamma_{\beta}} - 
2\int_{t_{0}}^{t}Wdt^{\prime}$ and 
\vskip 0,2 cm
$$M:= -{1\over 2}{({\dot W}/W\dot)} + {1\over 4}{({\dot W}/W)^{2}} - 
{9\over 4}{({\dot a}/a)^{2}} - {3\over 2}{({\dot a}/a\dot)} + 
{{\omega}^{2}} - {W^{2}}\eqno (5c)$$
\vskip 0,2 cm
where ${\omega}^{2}={k^{2}/a^{2}}+{m^2}$, ``$a$" is the scale factor of the 
universe and the dot is the cosmic time derivative. 

 Following  ref.[2] we can rewrite eqs. (4) in the form 
$$a_{\bf k} = e^{F} A_{\bf k} e^{-F}\eqno (6)$$
\vskip 0,2 cm
with the convenient notation
$$F={\sum _{{\bf k}=-\infty}}^{+\infty} F_{{\bf k},-{\bf k}}$$
$$F_{{\bf k},-{\bf k}}={1\over 2} r({\bf k},t)
[{e^{-2i\lambda ({\bf k},t)}}A_{\bf k}A_{-{\bf k}}-
{e^{2i\lambda ({\bf k},t)}{{A^{\dagger}}_{\bf k}}{A^{\dagger}}_{-{\bf k}}] 
+i{{\gamma}_{\alpha}}({\bf k},t)\{{A^{\dagger}}_{\bf k}},
A_{\bf k}\}$$ 
\vskip 0,2 cm
where $2\lambda  ({\bf k},t)=\gamma_{\alpha}({\bf k},t) + 
\gamma_{\beta}({\bf k},t)$.
\vskip 0,2 cm
 Defining also $\tilde F$ as:
$$\tilde F:= {\sum _{{{\bf k}^\prime}\not={\bf k}}}
F_{{{\bf k}^\prime},-{{\bf k}^\prime}}$$

 Using that $\tilde F$ commutes with $F_{-{\bf k},{\bf k}}$, 
$F_{{\bf k},-{\bf k}}$ and $A_{\bf k}$, we can rewrite eq.(6) in the form 
$$a_{\bf k}=exp (F_{-{\bf k},{\bf k}} + F_{{\bf k},-{\bf k}})A_{\bf k}
exp (-F_{-{\bf k},{\bf k}} - F_{{\bf k},-{\bf k}})\eqno (7)$$
\vskip 0,2 cm
Let us introduce the definitions  of  two-modes  
squeezed and single-mode rotation
operators as in ref.[7]. Using the commutation relation between $A_{\bf k}$ 
and ${A_{\bf k}}^{\dagger}$ and taking into account the isotropy of the space 
by means of the assumption
 
$$r(-{\bf k})=r({\bf k})$$
$${\gamma_{\alpha}} (-{\bf k})={\gamma_{\alpha}} ({\bf k})$$
$$\lambda (-{\bf k})=\lambda ({\bf k})$$ 
\vskip 0,2 cm
we can rewrite eq.(7) in the form 
$$a_{\bf k}(t)=S({r\over 2},{\lambda})R_{-{\bf k}}(-2{\gamma_{\alpha}})
S({r\over 2},{\lambda})R_{{\bf k}}(-2{\gamma_{\alpha}})A_{\bf k}$$
$${S^{\dagger}}({r\over 2},{\lambda})
{R^{\dagger}}_{-{\bf k}}(-2{\gamma_{\alpha}})
{S^{\dagger}}({r\over 2},{\lambda}){R^{\dagger}}_{{\bf k}}
(-2{\gamma_{\alpha}})\eqno (8)$$
\vskip 0,2 cm
where  
$$S({r\over 2},{\lambda})=\exp \{ {r({\bf k},t)\over 2}
[{e^{-2i\lambda ({\bf k},t)}}A_{\bf k}A_{-{\bf k}}-
e^{2i\lambda   ({\bf    k},t)}{{A^{\dagger}}_{\bf    k}}
{A^{\dagger}}_{-{\bf k}}]\}$$
\vskip 0,2 cm
is the two-mode squeezed operator, and 
$$R_{{\bf k}}(-2{\gamma_{\alpha}})=\exp \{2i{{\gamma}_{\alpha}}({\bf k},t)
{{A^{\dagger}}_{\bf k}} A_{\bf k}\}$$  
\vskip 0,2 cm
is the single-mode rotation operator.

 From ref.[7] we can use the relation
$$R(2\gamma)S(r/2,\lambda+\gamma)=S(r,\lambda)R(2\gamma)$$
$$R(-2\gamma){A_{\bf k}}R^{\dagger}(-2\gamma)={e^{-2i\gamma}}A_{\bf k}$$
Therefore we obtain 
$$a_{\bf k}(t)={e^{-2i\gamma}}S(r/2,\lambda)
S(r/2,\lambda+{{\gamma}_{\alpha}})A_{\bf k}
S^{\dagger}(r/2,\lambda+2{{\gamma}_{\alpha}})
S^{\dagger}(r/2,\lambda+{{\gamma}_{\alpha}})\eqno (9)$$

 In eq. (9) we can see that the deviation from 
the pure squeezing transformation is given by the 
function $\gamma_{\alpha}$. 

 We will study now the particular case 
$$\gamma_{\alpha}=0\eqno (10)$$
\vskip 0,2 cm
then we have 
$$a_{\bf k}(t)=S(r,\lambda){A_{\bf k}}{S^{\dagger}}(r,\lambda)\eqno (11)$$
\vskip 0,2 cm
with $S(r,\lambda)$ a two-mode squeezed operator. Following ref.[1] we can 
now expand eq.(11) in terms of only single-mode squeezed operators.  In order 
to do  that  we  can introduce the bosonic operators $b_{\bf k}$ and $b_{-{\bf
k}}$ by means of the definition 
$$A_{\bf k}={1\over {\sqrt 2}}({b_{sig({\bf k}){\bf k}}}
-i{sig({\bf k})}b_{-sig({\bf k}){\bf k}})\eqno (12) $$
\vskip 0,2 cm
where $sig({\bf k})$ is the sign function, defined by 
$$sig(-{\bf k})=-sig({\bf k})$$
and by  a  set  of  directions  which  are  choosen as positive.  We can
separate the states into the following disjoint sets $\{{\bf k}\}=
{{\Pi}_{\bf k}}\cup {{\Pi}_{-{\bf k}}}$, where 
$${{\Pi}_{\bf k}}=\{{\bf k}\ so\ that\ {\bf k}.{\bf u}>0\ when 
\ {\bf k}.{\bf  u}\not=0;$$
$$\  {\bf  k}.{\bf  v}>0\ when\ {\bf k}.{\bf v}\not=0\
and\ {\bf k}.{\bf u}=0;\ {\bf k}.({\bf v}\times{\bf u})>0\ when\ 
{\bf k}.{\bf u}={\bf k}.{\bf v}=0\}$$  
\vskip 0,2 cm
with ${\bf u}$ and ${\bf v}$ two arbitrary mutually perpendicular directions 
in the ${\bf k}$-space. Moreover
$${{\Pi}_{-{\bf k}}}=\{{\bf k}\ so\ that\ {\bf k}=-{{\bf k}^{\prime}},\ with\ 
{{\bf k}^{\prime}}\in {\Pi}_{\bf k}\}$$
\vskip 0,2 cm
Now we can define $sig({\bf k})=1$ if ${\bf k}\in {\Pi_{\bf k}}$ and -1 if 
${\bf k}\in {\Pi_{-{\bf k}}}$. In what follows we suppose that 
${\bf k}\in {\Pi_{\bf k}}$.

Then we can rewrite the two-mode squeezed operator as the product of 
single-mode operators:
$$S(r,\lambda)=S_{1+}(r,\lambda)S_{1-}(r,\lambda)\eqno (13)$$
\vskip 0,2 cm
with
\vskip 0,2 cm
$$S_{1\pm}=\exp [{r\over 2}
(e^{-2i\lambda}{b_{\pm{\bf k}}}^2 -
e^{2i\lambda}{{b^{\dagger}}_{\pm{\bf k}}}^2)]$$  
\vskip 0,2 cm

 Replacing eqs (12) and (13) in eq. (11) we obtain 
$$a_{\bf k} = {1\over {\sqrt 2}}({S_{1+}}{b_{\bf k}}{S^{\dagger}}_{1+} - 
i{S_{1-}}{b_{-{\bf k}}}{{S^{\dagger}}_{1-}})\eqno (14)$$

 Using now the procedure of ref.[1] we can factorize, the wave function 
$\Psi_{zk}$ in the space $(x_{+},x_{-})$ which corresponds respectively to 
$\Pi_{\bf k}$ and $\Pi_{-{\bf k}}$, in the form 
$${\Psi_{zk}}(x_{+}x_{-})= <x_{+}x_{-}|z_{k}>=
{{\Psi}^{1}}_{zk}(x_{+}){{\Psi}^{1}}_{zk}(x_{-})$$
\vskip 0,2cm

 It is not necessary to repeat the procedure of ref.[1], which finally 
permits us calculate the reduced density operator by means of 
$${\rho_{\bf k}}=\int dx_{+}dx_{-}|x_{+}x_{-}><x_{+}x_{-}|z_{k}>
<z_{k}|x_{+}x_{-}><x_{+}x_{-}|$$
\vskip 0,2cm
and the entropy using 
$$S=-Tr {\rho}\ln {\rho}$$
\vskip 0,2cm 
Therefore the 
entropy associated to the {\bf k}-mode 
can be calculated (as is shown in ref.[1]), by means of the 
approximate equation (valid when $n_{\bf k}>>1$):
$$S({\bf k})\simeq \ln n_{\bf k}\eqno (15)$$
\vskip 0,2 cm
But we  will  see  that  this  expression  gives  us  only the entropy for the
``in-out" process. We have not a dynamic equation for $S({\bf k})$, because 
the assumption given by eq.(10) is valid when $n_{\bf k}$ is a constant.

 We will analize now the physical consequences of the no rotation condition 
given by eq. (10). If ${\gamma}_{\alpha}$ is the the null function, eqs (5) 
turn to    
$$(1 + {\tanh r}\cos \Gamma)M = 0\eqno (16a)$$
\vskip 0,2 cm
$$M{\sin \Gamma} + 2W{\dot r} = 0\eqno (16b)$$
\vskip 0,2 cm
We have two options in order to satisfy eqs.(16): 
\vskip 0,2 cm
\item { a)} One possibility is by means of the following condition on the 
function $M$ (see eq. 5c), for any time $t \ge t_{1}$: 
$$M=0\eqno (17)$$
\vskip 0,2 cm
from eq. (16b) it turns out that the squeezing 
parameter $r_{\bf k}$ must be a 
constant in time. Therefore the mean 
value $n_{\bf k}$ of created particles  
between the times $t_{1}$ and $t$ is independent of time, because from 
eqs (4) we have 
$$n_{\bf k} = <0|{a^{\dagger}}_{\bf k}a_{\bf k}|0> = 
{\sinh }^{2}r_{\bf k}\eqno (18)$$

 Then, particles only can be created  if in equal number are annihilated. 

 The condition (17) is really a differential equation for the function $W$, 
which can be considered as the ``frequency" of the normal modes, in an 
expansion of the field as the one given by eq. (2).

 It is interesting to note that, when the process of interaction between 
the matter field and the geometry is considered as ``in-out", with 
$t_{1}=t_{in}$ and $t=t_{out}$, the conditions on $W(t_{out})$ and 
$\dot  W(t_{out})$,    that    in   the  literature  [8]  are  known  as  the
diagonalization of the  hamiltonian,  satisfy  automaticaly  eq.(17).   So a
process where the final result is a hamiltonian with the functional form of 
independent oscillators, produces squeezed states. 
\vskip 0,2 cm
\item { b)} The other possibility to satisfy eq. (16), is the requirement 
$$\cos \Gamma = - \coth r\eqno(19)$$

 Now performing the derivative  of  eq.    (19) and replacing in eq.  (16) we
obtain 
$${{\sinh}^2}r={M\over {2({{\dot \gamma}_{\beta}}-2W)W}}\eqno (20)$$
\vskip 0,2 cm
Putting eq.(20) in terms of the $\coth r$ and replacing in eq. (19), we get 
the following integro-differential equation for the function $W$ and 
the phase 
${\gamma}_{\beta}$: 
$${M[W]}^{1/2}\cos [{\gamma_{\beta}}-2{\int_{t_{0}}^{t}}Wdt^{\prime}]+
\{2W({{\dot \gamma}_{\beta}}-2W)+M[W]\}^{1/2}=0\eqno (21)$$
\vskip 0,2 cm

 From eqs (20) and (21)  we  can get a relation between the average number of
particles ``$n$"  and the function $W$. In order to do that we can rewrite 
eq. (20) in the form
$${{\dot \gamma}_{\beta}}={M\over {2nW}} + 2W$$
\vskip 0,2 cm
Integrating we have
$${\gamma}_{\beta} = {1\over 2} \int_{t_{0}}^{t} {M\over {nW}}dt^{\prime} + 
2\int_{t_{0}}^{t} Wdt^{\prime} + \mu \eqno (22)$$
\vskip 0,2 cm
with $\mu$ an integration constant. 

 Replacing now eq.(22) in eq.(21) and using the assumption that $M \ne 0$, 
in order to make the division by $M^{1/2}$, we get: 
$$\cos [\mu + {1\over 2}\int_{t_{0}}^{t}{M\over {nW}}]dt^{\prime} = 
-{[1+{1\over n}]}^{1/2}\eqno (23)$$
\vskip 0,2cm
Since  the module of the {\it cosine} must be greater than one and $n$ is a  
possitive integer,  eq.(23) only makes sense when $n=\infty$. 
Therefore the only non 
trivial possibility to solve eqs (16) is the one given in a). 
 In that case it is possible to calculate the entropy by means of 
eq. (15), as in ref. [1]. But, when the dynamics of the particle creation 
given by eq.(5) is taken into account, the transformation 
of the operators is given by eq. (9). 

 We will analize now, 
if it is still possible to apply the same approach to calculate the entropy, 
in the case when the dynamics, given by eqs (5), is not altered, 

The transformation between operators given by eq. (9) is not a pure 
squeezing one. However from eq.(9) we can obtain, using the unitarity of the 
operators; 
$$a_{\bf k}(t)|z_{\bf k}> = 0\eqno (24)$$
\vskip 0,2cm
with
$$|z_{\bf k}> := S(r/2, {\lambda} + {{\gamma}_{\alpha}})
S(r/2, {\lambda} + 2{{\gamma}_{\alpha}})|0>\eqno (25)$$
\vskip 0,2cm
then $|z>$ is a squeezed state, because it results of the application of two 
successive squeezing transformation.

    As  in  ref.  [1] we can introduce the
wave function $\Psi_{z{\bf k}}(x_{+},x_{-}) := <x_{+},x_{-}|z>$, which, from 
eq. (24), satisfies the equation 
$$a_{\bf k}{\Psi_{z{\bf k}}} = 0\eqno (26)$$
\vskip 0,2cm

 In order to make explicit eq. (26), we can take into account eq. (4a), 
replacing eq. (12) there, then we have the operator $a_{\bf k}(t)$ as a 
function of $b_{\bf k}$, $b_{-{\bf k}}$, and its hermitic conjugate:
$${\sqrt {2}}a_{\bf k}(t) =  e^{i\gamma_{\alpha}}\cosh {r} b_{\bf k}
+e^{i\gamma_{\beta}}\sinh {r} {b^{\dagger}}_{\bf k}-i(
e^{i\gamma_{\alpha}}\cosh {r} b_{-{\bf
k}}+e^{i\gamma_{\beta}}\sinh {r} {b^{\dagger}}_{-{\bf k}})\eqno (27)$$ 

 We  can now introduce  the following condition in eq.(27)
$${\gamma_{\beta}} = {\gamma_{\alpha}}\eqno (28)$$
\vskip 0,2cm
It was used that $\gamma_{\beta}$ does not appear either in eq. (5) or in the 
mean value of particle creation, therefore it is a ``gauge condition" that we 
can choose in order to obtain a Gaussian distribution for 
the wave function in full analogy with ref. [1]. 

 Taking into account the commutation relations $[b,b^{\dagger}]=1$ and 
$[{\hat x},{\hat  p}]=i$.    Calling  ${\hat  x}_{+}$  to  the superfluctuant
operator related with the operator $b_{\bf k}$ and ${\hat x}_{-}$ to the one 
corresponding to  $b_{-{\bf k}}$, we can write
 
$${b_{\bf k}}={{e^{i{\pi/2}}}\over {\sqrt {2}}} (x_{+} + 
{\partial _{x_{+}}})\ \ 
\ \ \ \ \ \ \ \ ,\ \ \ \ \ \ \ \ \ \ \ 
{b_{-{\bf k}}}={{e^{i{\pi/2}}}\over {\sqrt {2}}} (x_{-} + 
{\partial _{x_{-}}})\eqno (29)$$
\vskip 0,2cm
and analogously for the conjugated ones.
Replacing these expressions in eq.(27) it results  
$$i[\cosh r (x_{+} + {\partial _{x_{+}}}) - \sinh r (x_{+} - 
{\partial _{x_{+}}})]\Psi + [\cosh r (x_{-} + {\partial _{x_{-}}}) - 
\sinh r (x_{-} - {\partial _{x_{-}}})]\Psi = 0$$
\vskip 0,2cm

 Introducing the factorization 
$$\Psi ({x_{+}}{x_{-}}) := \Psi^{1}(x_{+})\Psi^{1}(x_{-})$$
\vskip 0,2cm

 For both variables the function $\Psi^{1}$ satisfies the equation 
$$\cosh r (x + \partial_{x})\Psi^{1} = \sinh r (x - \partial_{x})\Psi^{1}
\eqno (30)$$
\vskip 0,2cm

 Then $\Psi^{1}$ is a Gaussian distribution and all the calculation performed 
in ref.  [1] is available, so the entropy can be approximated by the logarithm  
of the mean value of the created particles.  
Therefore in a  way similar to that 
in  ref.    [9]  it  is  possible  to apply  this  formalism  to  calculate  the
cosmological entropy production.

 It is interesting to note that  we 
can get, for a large particle  creation  number, the same dependence with the
entropy, if we use the expression of the entropy corresponding to a bosonic 
gas in flat space time (see ref.[10]), but with the mean value of 
particle creation coming 
from the dynamics of the curved space time. 
\bigskip

{\bf ACKNOWLEDGEMENTS} 

 This work  was  supported  by  the  European  Community  DGXII  and  by  the
Departamento de F\'{\i}sica de la Facultad de Ciencias Exactas y Nat. de 
Buenos Aires. 
\bigskip

{\bf REFERENCES}
\vskip 0,2cm
\item{$[1]$} M.Gasperini  and  M..Giovannini, Phys. Lett. B 301 (1993) 334-338.
\item{$[2]$} L.Parker, Phys. Rev. {\bf 183}, $N^{o}$ 5, (1969), 1057-1068.
\item{$[3]$} L.P.Grishchuk and Y.V.Sidorov, Phys.  Rev.  D, {\bf 42}, $N^{o}$
10, (1990), 3413-3421.
\item{$[4]$}  M.Castagnino  and  C.Laciana, J.Math.Phys.  {\bf 29}, $N^{o}$2,
(1988), 460-475.
\item{$[5]$} N.D.Birrell  and  P.Davies, {\it Quantum Field Theory in Curved
Space} (Cambridge University Press, Cambridge, England, 1982). 
\item{$[6]$} C.E.Laciana, ``Thermal Conditions for Scalar Bosons in a 
Curved Space Time", Gen. Rel. and Grav., {\bf 28}, No. 8, (1966). 
\item{$[7]$} B.L.Schumaker,  Phys. Rep. 135 (1986) 317.
\item{$[8]$} A.A.Grib, S.G.Mamayev  and  V.M.Mostepanenko;   J.Phys.A:  Math.
Gen. {\bf 13} (1980) 2057-2065.
\item{$[9]$}  S.P.Kim and  S.W.Kim,  Phys.    Rev.    D,  {\bf  51},  (1995),
4254-4258.  
\item{$[10]$} L.D.Landau and E.M.Lifshitz, ``Statistical Physics", Vol.5. 
Pergamon Press (2nd Impression 1970), eq. 54.6, pag. 147.

\bye